\documentclass[twocolumn]{revtex4}[12pt]
\usepackage{graphicx}
\begin{document}
\title{Solitary waves and double layers in an adiabatic multi-component space plasma}
\author{M. G. Shah$^{a}$, M. M. Rahman$^{b}$, M. R. Hossen$^{c*}$, A. A. Mamun$^d$}
\address{$^{a}$Department of Physics, Hajee Mohammad Danesh Science and Technology University,
Dinajpur-5200, Bangladesh. \\
$^{b}$Department of EEE, European University of Bangladesh,
Shyamoli, Dhaka-1207, Bangladesh. \\
$^{c}$Department of General Educational Development, Daffodil
International University, Dhanmondi, Dhaka-1207, Bangladesh. \\
$^{d}$Department of Physics, Jahangirnagar University, Savar,
Dhaka-1342, Bangladesh. Email$^{*}$: rasel.plasma@gmail.com.}

\begin{abstract}
\noindent \textbf{Abstract}: The formation and propagation of
small amplitude Heavy-ion-acoustic (HIA) solitary waves and double
layers in an unmagnetized collisionless multi-component plasma
system consisting of superthermal electrons, Boltzmann
distributed light ions, and adiabatic positively charged inertial
heavy ions are theoretically investigated. The reductive
perturbation technique is employed to derive the modified
Korteweg-de Vries (mK-dV) and standard Gardner (SG) equations.
The solitary wave (SW) solution of mK-dV and SG equations as well
as Double Layers (DLs) solution of SG equation is studied for
analysis of higher-order nonlinearity. It is found that the
plasma system under consideration supports positive and negative
potential Gardner solitons but only positive potential mK-dV
solitons. In addition, it is shown that, the basic properties of
HIA mK-dV and Gardner solitons and DLs (viz. polarity, amplitude,
width, and phase speed) are incomparably influenced by the
adiabaticity effect of heavy ions and the superthermality effect
of electrons. The relevance of the present findings to the system
of space plasmas as well as to the system of researchers interest is specified.
\end{abstract}

\maketitle
\section{Introduction}
The physics of astrophysical plasmas plays the part of most
expeditiously growing branches of plasma physics, because such
plasmas are omnipresent in space environments
\cite{Rees1983,Miller1987,Michel1991,Havnes1992,Whipple1992}. The
ion-acoustic (IA) waves have a crucial role for learning the
linear and nonlinear characteristics of various astrophysical
plasmas. This type of wave modes are nothing but a low-frequency
longitudinal plasma density oscillations which was first
anticipated in 1929 by Tonks and Langmuir who defined the phase
velocity for isothermal changes and frequencies well below the
ion plasma frequency \cite{Tonks1929}. In 1933, Revans studied
the IA waves experimentally and observed the presence of
electrical oscillations within an electrical discharges through
gases \cite{Revans1933}.

The study of a small population of excess energetic or
superthermal electrons within a space plasma system with strong
electric fields has proven to be a rich topic with many
surprising new phenomena discovered in recent and past years.
Astrophysical plasmas are usually hot and collisionless and their
characteristics are maintained by the collective wave-particle
interactions preferably than individual particle-particle
interactions. A number of Satellite or experimental observations
have uncovered the presence of excess energetic or superthermal
electrons in space and laboratory plasma environments
\cite{Vasyliunas1968,Formisano1973,Scudder1981,Feldman1982,Marsch1982,Collier1996,Mori2003,Papp2015,Zeng2015}.
The plasma system that contains highly energetic particles with
energies greater than thermal energies is studied by using the
kappa velocity distribution
\cite{Vasyliunas1968,Baluku2008,Baluku2010}. Notably, high energy
particles exist due to the effect of external forces acting on
the space environment plasmas. Gurevich \textit{et al.}
\cite{Gurevich1995} investigated the behaviour of fast or
superthermal electrons by deriving the kinetic equation and
demonstrated the strong influence of the density profile on the
diffusion of fast particles. Lu \textit{et al.} \cite{Lu2011}
studied the presence of superthermal electrons in the space plasma
and observed the variation of indices of the power law
distribution in the magnetosheath. For modeling aforesaid plasmas,
kappa velocity distribution is applicable because Boltzmann
distribution perhaps inappropriate for explaining the long-range
interactions. The generalized three dimensional kappa
distribution function \cite{Summer1991} is mathematically
represented as,

$$F_k(v)=\frac{\Gamma(k+1)}{(\pi k
w)^{3/2}\Gamma(k-1/2)}(1+\frac{v^2}{kw^2})^{-(k+1)}$$

Where $F_k$ represents the kappa distribution function, $\Gamma$
is the gamma function, $w$ being the most probable speed of the
energetic particles, given by
$w=[(2k-3/k)^{1/2}(k_{B}T/m)^{1/2}]$, with T shows the
characteristic kinetic temperature and $w$ is related to the
thermal speed $V_{t}=(k_{B}T/m)^{1/2}$ and, the parameter $k$
symbolizes the spectral index \cite{Cattaert2007} which defines
the strength of the superthermality. The range of this parameter
is $3/2<k<\infty$ \cite{Alam2013}. In the limit
$k\rightarrow\infty$ \cite{Basu2008,Baluku2012}, the kappa
distribution function lessens to the well-known Boltzmann
distribution.

The study of plasmas with impurities (heavy ions or charged dust
grains) has been attracted a great deal of interest among the
researchers
\cite{Hines1957,Rao1990,popel1995a,popel1996,Shukla2002,Losseva2009}.
Rao \textit{et al.} \cite{Rao1990} considered a dusty plasma
system containing dust particles and investigated the linear and
nonlinear wave phenomena of dust-acoustic waves. In our present
work, the theoretical model is assumed to be composed of inertial
heavy ions which is millions to billions times lighter than that
of the dust particles. The nonlinear behavior of ion acoustic
waves for plasmas containing impurity ions or negatively charged
dust particles is examined by Popel and Yu \cite{popel1995a} since
positively charged inertial heavy ions are taken into
consideration in our present model. The HIA waves differs from
conventional ion acoustic waves in impurity (dust grains)
containing plasmas by the following points: i) in
Dust-ion-acoustic (DIA) waves the inertia is given by the light
ions mass whereas in HIA waves inertia is provided by the heavy
ion mass; ii) in HIA waves mobile heavy ions provide the
necessary inertia whereas in DIA waves static dust grains
participate only in maintaining the equilibrium charge neutrality
condition; iii) the frequency of the DIA  waves is much greater
than that of the HIA waves.

A significant number of attempts have been taken to study the
characteristics of plasma system that usually contains
ions/heavy-ions or electrons
\cite{mrHossen2014a,mrHossen2014b,mrHossen2014c,mrHossen2014d,mrHossen2014e,BHosen2016a,BHosen2016b,Sayed2007,Ema2015a,Ema2015b,Rahman2007,Mamun2008,Mahmood2008,Fatema2008,Hossen2014e,Hossen2014f,Hossen2014g,Hossen2014h}.
Mahmood \textit{et al.} \cite{Mahmood2008} studied the IA waves
in a multi-component plasma system consisting of adiabatically
heated ions and observed that the amplitude of electron density
humps decreases with the increase of hot ion temperatures. A
degenerate quantum plasma system is thought out by Hossen
\textit{et al.}
\cite{Hossen2014e,Hossen2014f,Hossen2014g,Hossen2014h} who
studied the roles of heavy ions on HIA waves. In former time,
Hines considered a multi-ion plasma system in the ionosphere and
studied the effect of heavy ions on the propagation of radio waves
\cite{Hines1957}. Mamun \textit{et al.} \cite{Mamun2008} observed
the combined effects of adiabatic electrons and negatively
charged static dust that how they modify the basic properties of
the DIA K-dV solitons by studying a dusty plasma system. A lot of
investigations have also been made to study the energetic
particles by a number of plasma scientists. Barbosa \textit{et
al.} \cite{Barbosa1980} considered a model for the generation of
banded electrostatic emissions by superthermal electrons
employing a power law form which can apply to Jupiter's
magnetosphere and concentrated on instability in the upper hybrid
and lower harmonic bands. Basu \cite{Basu2008} addressed a plasma
system with kappa distributed plasma particles in order to
explain the distinguishing features of the kappa distributions.
Using kinetic theoretical approach, the basic features of electron
acoustic waves in a plasma system whose electrons are two-kappa
distributed are studied by Baluku \textit{et al.}
\cite{Baluku2011} . Mehran \cite{Mehran2012} considered an
unmagnetized plasma including cool ions and hot ions with kappa
distributed electrons and investigated the basic properties of IA
waves where the suprathermality effects play vital roles. Later,
Baluku \textit{et al.} \cite{Baluku2012} studied the features of
IA solitons in a plasma system with both electron components are
kappa-distributed found in Saturn's magnetosphere. Eslami
\textit{et al.} \cite{Eslami2011} thought out an unmagnetized
plasma system containing warm adiabatic ions, superthermal
electrons, and thermal positrons and found the effects of the
spectral index $k$ on the IA waves.

Therefore, the effects of superthermal electrons and adiabatic
heavy ions on the propagaton of HIA waves in EI plasmas have been
investigated. We have studied the basic properties of small
amplitude HIA waves by using the reductive perturbation method in
concerning plasma system. To the best of our knowledge no
theoretical investigations have been made to investigate the HIA
solitary waves in EI plasmas comprising of superthermal electrons,
Boltzmann distributed light ions, and adiabatic positively charged
inertial heavy ions. Thus, our present attempt is to investigate
the contribution of superthermality of electrons and adiabaticity
of heavy ions on the nonlinear propagation of HIA waves.

The remainder of this paper is arranged as follows: The basic
governing equations are provided in sect. II. The mK-dV and SG
equations are derived in sects. III and IV, and their solitary
wave solutions are analyzed in sect. V, respectively. DL solution
of SG equation is analyzed in sect. VI. Finally, a brief results
and discussion is presented in sect. VII.


\section{The basic governing equations}

\label{noi2} The propagation of HIA waves in an unmagnetized,
collisionless plasma system containing superthermal electrons,
Boltzmann distributed light ions, and adiabatic positively
charged inertial heavy ions has been considered. At equilibrium,
we have $Z_in_{i0}+Z_hn_{h0} = n_{e0}$, where  $n_{i0}, n_{h0}$,
and $n_{e0}$ are the unperturbed light ion, heavy ion, and
electron number density, $Z_i$ is the ion number state, and $Z_h$
is the number of light ions residing on the heavy ion's surface.

The dynamics of HIA waves in such an adiabatic plasma system is
governed by the following equations:

\begin{eqnarray}
&&\hspace*{10mm}\frac{\partial n_h}{\partial t} +
\frac{\partial}{\partial x}(n_hu_h) = 0,
\label{be1}\\
&&\hspace*{10mm}\frac{\partial u_h}{\partial t} + u_h
\frac{\partial u_h} {\partial x} + \frac{\partial \psi} {\partial
x} + \frac{\sigma}{n_h} \frac{\partial p_h}{\partial x}=0,
\label{be2}\\
&&\hspace*{10mm}\frac{\partial p_h}{\partial t} + u_h
\frac{\partial p_h} {\partial x} + \gamma_h p_h \frac{\partial
u_h}{\partial x}=0,
\label{be3}\\
&&\hspace*{10mm}\frac{\partial^2 \psi}{\partial x^2}=-n_h+\beta
n_e-\alpha n_i,\label{be4}
\end{eqnarray}

\noindent where the number density of the plasma species $(n_j)$
(j= h, i, e; h for heavy ion, i for light ion, e for electron),
heavy ion number density $(n_h)$, heavy ion thermal pressure
$(p_h)$, heavy ion fluid speed $(u_h)$, and electrostatic
potential $(\psi)$ are normalized by ion/electron equilibrium
number density $n_{j0}$, heavy ion number density at equilibrium
multiplied by number of electrons residing on the heavy ion
$Z_hn_{h0}$, $Z_hn_{h0}T_h$, effective heavy ion acoustic
velocity $C_h=\sqrt{T_e/m_h}$, and the quantity $T_e/e$. The
space and time variables are normalized by the Debye radius
$\lambda_{D}=\sqrt{T_e/4 \pi Z_hn_{h0}e^2}$ and the reciprocal
heavy ion plasma frequency ${\omega_{h}}^{-1}= (4 \pi e^2
Z_hn_{h0}/m_h)^{1/2}$, e is the magnitude of the charge of
electron. Furthermore, $\sigma=(T_h/T_e)$,
$\alpha=(Z_in_{i0}/Z_hn_{h0})$, and $\beta=(n_{e0}/Z_hn_{h0})$.

It is noted here that for an isothermal process $\gamma_s=1$ and
$p_s=n_s$ with constant $T_s$ $(i.e., T_s=T_{s0})$, where $s$ is
the concerning plasma species taken as adiabatic in the
corresponding model. For adiabatic ion fluid, $\gamma_s=3$ is
taken into account\cite{Mamun2008}.

For Modelling the effects of superthermal electrons we have
considered the following kappa distribution
\begin{eqnarray}
&&\hspace*{5mm}n_e=n_{e0}[1-\frac{e\psi}{{T_e}(k-\frac{3}{2})}]exp(-k+\frac{1}{2}),\label{1}\
\end{eqnarray}

\noindent here, the superthermal parameter k stands for kappa
distribution. It is notable that the range of k is $1.6\sim2.2$
\cite{Eslami2012,Alam2014}.

The well-known Maxwell-Boltzmann distribution has been considered
for light ions. In this case, we have obtained the following
Maxwellian light ion number density.

\begin{eqnarray}
&&\hspace*{5mm}n_i=n_{i0}exp(\frac{-e\psi}{T_i}),\label{2}
\end{eqnarray}

\noindent where, $n_e$ $(n_i)$ is the number density of the
perturbed electrons (light ions) and $T_e$ $(T_i)$ is the
temperature of electrons (light ions), respectively.


\section{Derivation of the modified K-dV equation}
We have examined the electrostatic perturbations propagating in
an unmagnetized collisionless  plasma system due to the effect of
dispersion. By considering higher order term, we have derived
mK-dV equation employing reductive perturbation method using Eqs.
(\ref{be1}) - (\ref{be4}), a set of stretched coordinates
\cite{Maxon1974,Shah2015a,Shah2015b,Rahman2014} has introduced for
the mK-dV equation as
\begin{eqnarray}
&&\hspace*{5mm}\eta ={\epsilon}(x - V_pt),
\label{sa}\\
&&\hspace*{5mm}T={\epsilon}^{3}t \label{sb}
\end{eqnarray}

\noindent where $V_p$ is the wave phase speed ($\omega/k$ with
$\omega$ being angular frequency and $k$ being the wave number of
the perturbation mode), and $\epsilon$ is a smallness parameter
measuring the weakness of the dispersion ($0<\epsilon<1$). Now, we
have expanded $n_j$, $u_h$, $p_h$ and $\psi$ in power series of
$\epsilon$ in the following way,

\begin{eqnarray}
&&\hspace*{5mm}n_j=1+\epsilon
n_j^{(1)}+\epsilon^{2}n_j^{(2)}+\epsilon^{3}n_j^{(3)}+ \cdot \cdot \cdot, \label{6a}\\
&&\hspace*{5mm}u_h=0+\epsilon u_h^{(1)}+\epsilon^{2}u_h^{(2)}+\epsilon^{3}u_h^{(3)}+\cdot \cdot \cdot,\label{6c}\\
&&\hspace*{5mm}p_h=1+\epsilon p_h^{(1)}+\epsilon^{2}p_h^{(2)}+\epsilon^{3}p_h^{(3)}+\cdot\cdot\cdot,\label{6d}\\
&&\hspace*{5mm}\psi=0+\epsilon\psi^{(1)}+\epsilon^{2}\psi^{(2)}+\epsilon^{3}\psi^{(3)}+\cdot\cdot
\cdot.\label{6e}
\end{eqnarray}

\noindent By substituting the values of $\eta$, $T$, and
expansions (\ref{6a})-(\ref{6e}) into (\ref{be1}) - (\ref{be4})
and taking the coefficient of $\epsilon^{2}$ from equations
(\ref{be1})-(\ref{be3}), and $\epsilon$ from (\ref{be4}), we have
$n_j^{(1)}=u_j^{(1)}/V_p$, $u_h^{(1)}=
V_p\psi^{(1)}/V_p^2-\gamma\sigma$, $n_h^{(1)}=
\psi^{(1)}/V_p^2-\gamma\sigma$,
$p_h^{(1)}=\gamma\psi^{(1)}/(V_p^2-\gamma\sigma)$, and
$$V_p=\sqrt{\frac{2k-3}{\alpha\sigma_i(2k-3)+\beta(2k-1)}+\gamma\sigma}$$
where $\sigma_i=T_e/T_i$ and $V_p$ represents the dispersion
relation for the HIA type electrostatic waves in an EI plasma
under consideration.

Taking the coefficient of $\epsilon^{3}$, we have obtained a set
of equations, which, after using the values of $n_{h}^{(1)}$,
$u_{h}^{(1)}$, and $p_{h}^{(1)}$, can be simplified as

\begin{eqnarray}
&&\hspace*{-1mm}n_{h}^{(2)}=\frac{3V_p^{2}-2\sigma\gamma+\sigma\gamma^2}{2(V_p^{2}
-\sigma\gamma)^{3}}(\psi^{(1)})^{2}+\frac{\psi^{(2)}}{V_p^{2}-\sigma\gamma},\label{SL1}\\
&&\hspace*{-1mm}u_{h}^{(2)}=\frac{V_p^{3}+V_p\sigma\gamma^2}{2(V_p^{2}
-\sigma\gamma)^{3}}(\psi^{(1)})^{2}+\frac{V_p\psi^{(2)}}{V_p^{2}-\sigma\gamma},\label{SL2}\\
&&\hspace*{-1mm}p_{h}^{(2)}=\frac{\gamma(\psi^{(1)})^{2}}{2(V_p^{2}
-\sigma\gamma)^{2}}+\frac{\gamma
u_{h}^{(2)}}{V_p}+\frac{\gamma^2(\psi^{(1)})^{2}}{2(V_p^{2}-\sigma\gamma)^2},
\label{SL3}\\
&&\hspace*{-1mm}{\rho^{(2)}}=\frac{1}{2}A(\psi^{(1)})^{2},\label{SL3b}
\end{eqnarray}

\noindent where

\begin{eqnarray}
&&\hspace*{-1mm}A=[\frac{3V_p^{2}-2\sigma\gamma+\sigma\gamma^2}{(V_p^{2}
-\sigma\gamma)^{3}}+\beta\frac{1-4k^2}{(2k-3)^2}+\alpha{\sigma_i}^{2}]
\end{eqnarray}

To the next higher order of $\epsilon$, we obtain a set of
equations

\begin{eqnarray}
&&\hspace*{-2mm}\frac{\partial n_{h}^{(1)}}{\partial
T}-V_{p}\frac{\partial n_{h}^{(3)}}{\partial \eta}
+\frac{\partial u_{h}^{(3)}}{\partial
\eta}+\frac{\partial}{\partial
\eta}[u_{h}^{(2)}n_{h}^{(1)}\nonumber\\
&&\hspace*{35mm}+n_{h}^{(2)}u_{h}^{(1)}]=0,
\label{SL4}\\
&&\hspace*{-2mm}\frac{\partial u_{h}^{(1)}}{\partial
T}-V_p\frac{\partial u_{h}^{(3)}}{\partial \eta}
+\frac{\partial}{\partial\eta}[u_{h}^{(1)}u_{h}^{(2)}]+\frac{\partial
\psi^{(3)}}{\partial \eta}\nonumber\\
&&\hspace*{18mm}+\sigma\frac{\partial p_{h}^{(3)}}{\partial
\eta}-\sigma n_{h}^{(1)}\frac{\partial p_{h}^{(2)}}{\partial
\eta} = 0,
\label{SL5}\\
&&\hspace*{-2mm}\frac{\partial p_{h}^{(1)}}{\partial
T}-V_p\frac{\partial p_{h}^{(3)}}{\partial \eta}
+u_{h}^{(1)}\frac{\partial p_{h}^{(2)}}{\partial
\eta}+u_{h}^{(2)}\frac{\partial p_{h}^{(1)}}{\partial
\eta}\nonumber\\
&&\hspace*{2mm}+\gamma\frac{\partial u_{h}^{(3)}}{\partial
\eta}+\gamma p_{h}^{(1)}\frac{\partial u_{h}^{(2)}}{\partial
\eta}+\gamma p_{h}^{(2)}\frac{\partial u_{h}^{(1)}}{\partial
\eta} = 0,
\label{SL6}\\
&&\hspace*{-2mm}\frac{\partial^2 \psi^{(1)}}{\partial
\eta^2}=-n_h^{(3)}-
\beta\left[\frac{(2k+3)(1-4k^2)}{6(2k-3)^3}(\psi^{(1)})^3\right.\nonumber\\
&&\hspace*{10mm}\left.+\frac{(1-4k^2)}{(2k-3)^2}\psi^{(1)}\psi^{(2)}+\frac{(1-2k)}{2k-3}\psi^{(3)}\right] \nonumber\\
&&\hspace*{8mm}+\alpha\left[{\sigma_i}{\psi^{(3)}}-{\sigma_i}^2\psi^{(1)}\psi^{(2)}+\frac{1}{6}{\sigma_i}^3({\psi^{(1)}})^3\right]
\label{SL8}\
\end{eqnarray}
\noindent Now combining Eqs. (\ref{SL4}) - (\ref{SL8}) and using
the values of $n_{h}^{(2)}$, $u_{h}^{(2)}$, $p_h^{(2)}$ and
$\rho^{(2)}$, we obtain an equation of the form
\begin{eqnarray}
&&\hspace*{-2mm}\frac{\partial\psi^{(1)}}{\partial T} +
\alpha_{1}\alpha_{2}({\psi^{(1)}})^2\frac{\partial
\psi^{(1)}}{\partial \eta} + \alpha_{2}\frac{\partial^3
\psi^{(1)}}{\partial \eta^3}= 0, \label{mK-dV}
\end{eqnarray}
\noindent where the value $\alpha_{1}$ and $\alpha_{2}$ is given
by
\begin{eqnarray}
&&\hspace*{-7mm}\alpha_{1}=\frac{15V_p^{4}+9V_p^{2}\sigma\gamma^2-17V_p^{2}\sigma\gamma-5\sigma^2\gamma^3}{2(V_p^2-\sigma\gamma)^{5}}\nonumber\\
&&\hspace*{5mm}+\frac{3\sigma^2\gamma^4+6\sigma^2
\gamma^2}{2(V_p^2-\sigma\gamma)^{5}}+\frac{\sigma\gamma^2+\sigma\gamma^3}{2(V_p^2-\sigma\gamma)^{4}}
\nonumber\\
&&\hspace*{5mm}+\frac{\beta(2k+3)(1-4k^2)}{2(2k-3)^3}-\frac{\alpha}{2}\sigma_{i}^3,\label{mK-Va}\\
&&\hspace*{-7mm}\alpha_{2}=\frac{{(V_p^2-\sigma\gamma)}^2}{2V_p},\label{mK-Vb}
\end{eqnarray}

In order to trace the influence of different plasma parameters on
the propagation of HIA waves in our considered plasma system, we
have derived the mK-dV equation (\ref{mK-dV}). The stationary
solitary wave solution of standard mK-dV equation is obtained by
considering a frame $\xi=\eta-u_{0} T$ (moving with speed $u_{0}$)
and the solution is,

\begin{eqnarray}
\nonumber\\
&&\hspace*{2mm}{\rm \psi^{(1)}}=\rm \psi_m{\rm sech
}(\frac{\xi}{\delta}), \label{solK-dV}
\end{eqnarray}

\noindent where the amplitude, $\psi_m=
\sqrt{6u_{0}/\alpha_{1}\alpha_{2}}$, and the width, $\delta=
\psi_m \sqrt{\gamma_1},{\gamma_1}=\alpha_{1}/6$ and $u_{0}$ is
the plasma species speed at equilibrium. The amplitude and width
variation of mK-dV solitons are nearly valid around critical
value $(k=k_c)$. The mK-dV equation has a SW solution around
$k=k_c$, but not any DLs solution. Therefore, we proceed into
next higher-order nonlinear equation known as SG equation,
because the SG equation supports both SWs and DLs just at the
critical value obtained from $A=0$.

\begin{figure}[t!]
\centerline{\includegraphics[width=6.5cm]{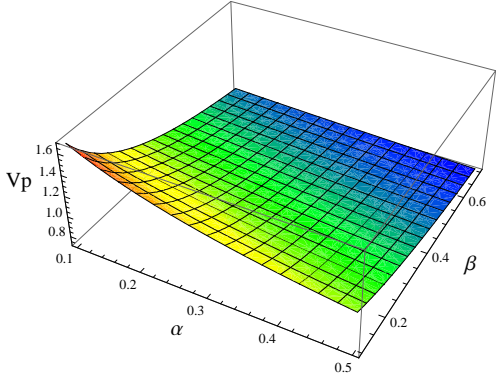}} \caption{The
variation of phase speed $V_p$ with $\alpha$ and $\beta$. The
other parameters are fixed at $k=2.3$, $\sigma=0.03$,
$\sigma_i=1.4$, and $U_{0}=0.01$.} \label{Fig.1}
\end{figure}
\section{Derivation of SG Equation}
\noindent It is obvious from Eq. (\ref{SL3b}) that $A=0$ since
$\psi^{(1)}\neq0$. We also found that $A=0$ around its critical
value $k=k_{c}$. We then assume $A=A_{0}$ when $k\neq k_{c}$ and
$k$ is close to $k_{c}$ that $\mid k=k_{c} \mid=\epsilon$ and we
can represent $A=0$ as
\begin{eqnarray}
\nonumber\\
&&\hspace*{2mm}A_0\simeq s\left(\frac{\partial A}{\partial
k}\right)_{k=k_{c}}|k-k_{c}|=sC_{1}\epsilon, \label{a33}
\end{eqnarray}
where $|k-k_{c}|$ is a small and dimensionless parameter, and can
be taken as the expansion parameter $\epsilon $, i.e.
$|k-k_{c}|\simeq \epsilon $, and $s=1$ for $k<k_{c}$ and $s=-1$
for $k>k_{c}$. $C_{1}$ is a constant depending on the
superthermality parameter $k$, and is given by
\begin{eqnarray}
\nonumber\\
&&\hspace*{3mm}C_{1}=\frac{1-4k^2}{(2k-3)^2}.\label{aa33}
\end{eqnarray}
So, $\rho^{(2)}$ can be expressed as
\begin{eqnarray}
\nonumber\\
&&\hspace*{3mm}\epsilon^2\rho^{(2)}\simeq
-\epsilon^3\frac{1}{2}C_{1}s({\psi^{(1)}})^2, \label{a34}
\end{eqnarray}

\begin{figure}[t!]
\centerline{\includegraphics[width=6.5cm]{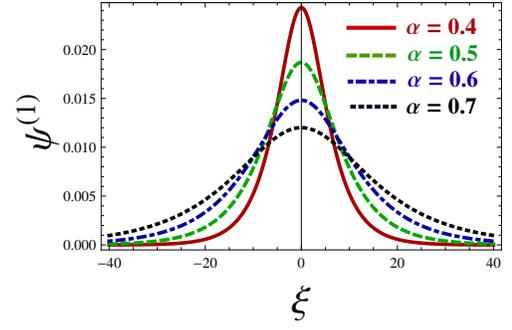}} \caption{The
variation of the positive potential mK-dV solitons with $\alpha$.
The other parameters are fixed at $k=2.3$, $\sigma=0.03$,
$\beta=0.4$, $\sigma_i=1.4$, and $U_{0}=0.01$.} \label{Fig.2}
\end{figure}

\noindent which, therefore, must be included in the third order
Poisson's equation. To the next higher order in $\epsilon^3$, we
obtain the following equation
\begin{eqnarray}
&&\hspace{-7mm}\frac{\partial^2\psi^{(1)}}{\partial
\eta^2}+\frac{1}{2}sC_{1}({\psi^{(1)}})^2-n_{h}^{(3)}-\beta\left[\frac{(2k+3)(1-4k^2)}{(2k-3)^3}\right.\nonumber\\
&&\hspace*{-5mm}\left.({\psi^{(1)}})^3+\frac{(1-4k^2)}{(2k-3)^3}\psi^{(1)}\psi^{(2)}+\frac{(1-2k)}{(2k-3)}\psi^{(3)}\right]\nonumber\\
&&\hspace*{-2mm}-\alpha\left[\sigma_{i}\psi^{(3)}-\sigma_{i}^2\psi^{(1)}\psi^{(2)}
+\frac{1}{6}\sigma_{i}^3({\psi^{(1)}})^3\right]=0.\label{a35}
\end{eqnarray}
After simplification, we can write from Eq. (\ref{a35})
\begin{eqnarray}
\nonumber\\
&&\hspace*{-7mm}\frac{\partial\psi^{(1)}}{\partial
T}+sC_{1}\alpha_{2}\psi^{(1)}\frac{\partial \psi^{(1)}}{\partial
\eta} +\alpha_{1}\alpha_{2}({\psi^{(1)}})^2\frac{\partial
\psi^{(1)}}{\partial \eta}\nonumber\\
&&\hspace*{38mm}+\alpha_{2}\frac{\partial^3 \psi^{(1)}}{\partial
\eta^3}=0. \label{sGS}
\end{eqnarray}
Equation (\ref{sGS}) is known as SG equation. It is also called
mixed mK-dV equation \cite{Lee2009}. It supports both the SWs and
DLs solutions since it contains both $\psi$-term of
korteweg-de-Vries (K-dV) and $\psi^2$-term of mK-dV equation. The
Gardner equation derived here is valid for $k\simeq k_{c}$.
Figures 5-7 show the variation of the amplitude of positive and
negative GSs with $k$ and number density ratio $\beta$
respectively.

\begin{figure}[t!]
\centerline{\includegraphics[width=6.5cm]{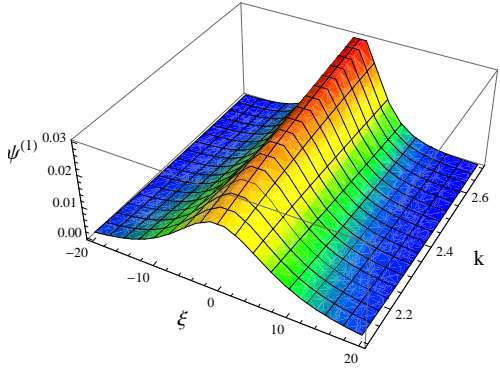}} \caption{The
variation of the positive potential mK-dV solitons for different
values of spectral index $k$. The other parameters are fixed at
$\alpha=0.4$, $\sigma=0.03$, $\beta=0.4$, $\sigma_i=1.4$, and
$U_{0}=0.01$.} \label{Fig.3}
\end{figure}


\section{SW Solution of SG Equation}
\noindent The SW solution of SG equation is given by the following
equation \cite{Alam2014}:
\begin{eqnarray}
&&\hspace*{2mm}\psi^{(1)}=\left[\frac{1}{\psi_{m2}}-\left(\frac{1}{\psi_{m2}}
-\frac{1}{\psi_{m1}}\right)\cosh^2\left(\frac{\xi}{\Delta}\right)\right]^{-1},
\label{e47}
\end{eqnarray}

\noindent where

\begin{eqnarray}
&&\hspace*{2mm}\psi_{m1,2}=\psi_m\left[1\mp\sqrt{1+\frac{U_0}{V_0}}\right],\label{e45}\\
&&\hspace*{2mm}U_{0}=\frac{sC_{1}B}{3}\psi_{m1,2}+\frac{\alpha_1B}{6}\psi_{m1,2}^2,\label{a44}\\
&&\hspace*{2mm}V_0=\frac{C_{1}^2s^2B}{6\alpha_1},\label{ab44}\\
&&\hspace*{2mm}\psi_m=\frac{-C_{1}s}{\alpha_1},\label{ab45}\\
&&\hspace*{2mm}\Delta=\frac{2}{\sqrt{-\gamma\psi_{m1}\psi_{m2}}}.\label{a48}\\
&&\hspace*{2mm}\gamma=\frac{\alpha_1}{6}.\label{aa48}
\end{eqnarray}

\begin{figure}[t!]
\centerline{\includegraphics[width=6.5cm]{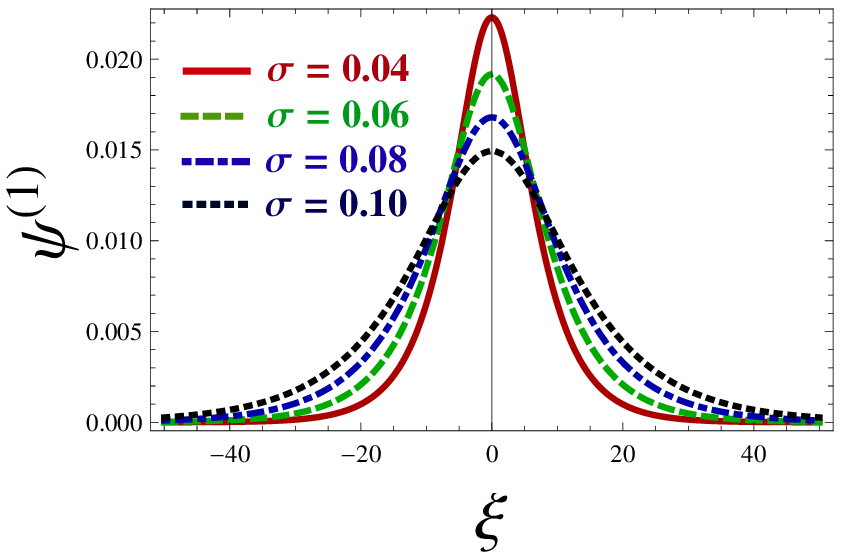}} \caption{The
variation of the positive potential mK-dV solitons with $\sigma$.
The other parameters are fixed at $k=2.3$, $\alpha=0.4$,
$\sigma_i=1.4$, $\beta=0.4$, and $U_{0}=0.01$.} \label{Fig.4}
\end{figure}

\section{DL Solution of SG Equation}
\noindent The DL solution of Eq. (\ref{sGS}) is given by
\begin{eqnarray}
&&\hspace*{2mm}\psi^{(1)}=\frac{\psi_m}{2}\left[1+\tanh \left
(\frac{\xi}{\Delta} \right)\right],\label{SolDL}
\end{eqnarray}
with
\begin{eqnarray}
&&\hspace*{2mm}U_0=-\frac{s^2B}{6\alpha_1},\label{2e50}\\
&&\hspace*{2mm}\psi_m=\frac{6U_0}{sB},\label{251}\\
&&\hspace*{2mm}\Delta=\frac{2}{\psi_m\sqrt{-\gamma}}, \label{254}
\end{eqnarray}

\noindent where $\gamma=\alpha_1/6$ and $\psi_m$ ($\Delta$) is the
DL height (thickness). This clearly indicates that Eq.
(\ref{SolDL}) represents a DL solution if and only if $\gamma<0$,
i.e., $\alpha_1<0$. When $\alpha_1B=0$, then, in this case, we
can find the value of the parameter $k$ as the critical one
($k=k_{d}\approx1.29$) for a set of plasma parameters viz.
$\sigma_i=1.6$, $\beta=0.7$, and $U_{0}=0.01$. For $k_{d}<1.29$,
$\alpha_1<0$, DL does not form. Therefore, positive potential DLs
exist for $k_{d}>1.29$ and $s=1$ in our considered plasma system.
Fig. 8 indicates the variation of the amplitude of positive DLs
with $\alpha$.

\begin{figure}[t!]
\centerline{\includegraphics[width=6.5cm]{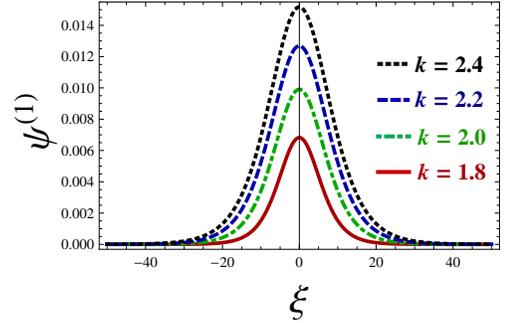}} \caption{The
variation of the positive potential GS solitons with $k$. The
other parameters are fixed at $\alpha=0.4$, $\sigma_i=1.4$,
$\beta=0.4$, and $U_{0}=0.01$.} \label{Fig.5}
\end{figure}

\begin{figure}[t!]
\centerline{\includegraphics[width=6.5cm]{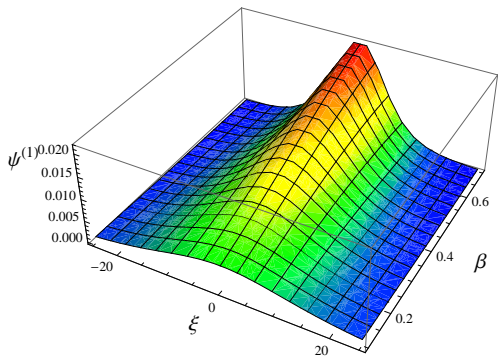}} \caption{The
variation of the positive potential GS solitons for different
values of $\beta$. The other parameters are fixed at $k=2.3$,
$\alpha=0.4$, $\sigma_i=1.4$, and $U_{0}=0.01$.} \label{Fig.6}
\end{figure}

\section{Results and Discussion}

In our present investigation, the small amplitude HIA solitary
waves rely on plasma parameters, namely the ion to heavy ion
number density ratio $\alpha$, the electron to heavy ion number
density $\beta$ and th temperature ratio $\sigma$. From the
expression of $V_p$ it is clear that the phase speed decreases by
the plasma parameters $\alpha$ and $\beta$. Figures 1-8 show how
the basic features of the HIA waves are significantly varied in
the presence of superthermal electrons, Boltzmann distributed
light ions, and adiabatic positively charged inertial heavy ions.
The phase speed $V_p$ has been changed moderately due to the
effects of $\alpha$ and $\beta$ which are shown in Fig. 1. The
amplitude of the HIA solitary waves decreases with the increasing
of $\alpha$ and $\beta$ due to the increasing of inertia (see
Figs. 2 and 4). The spectral index $k$ has a vital role on the
forming of solitary structures. For small values of $k$ the
superthermal electrons in the tail of velocity distribution
function increases and, vice versa. In case of mK-dV solitons
only positive potentials have been found which is shown in Figs.
2-4 for $(k_c>1.48)$ but below this critical value no profiles
have been seen. The superthermality effects are presented in
Figs. 5 and 7 where the positive and negative potentials of
Gardner solitons have been found for $(k_c>1.48)$ and
$(k_c<1.48)$, respectively. DL solution of equation has been
analyzed by considering a new set of plasma parameters
$\sigma_i=1.6$, $\beta=0.7$, and $U_{0}=0.01$. The positive
potential DLs has been found for $(k_d>1.29)$ depicted in Fig. 8.
Eventually, the results that we have found in our present
investigation can be summarized as follows:

\begin{enumerate}

\item{The plasma system under consideration supports finite but small
amplitude HIA solitary structures whose fundamental
characteristics (viz. polarity, amplitude, phase speed, etc.)
have been found distinctly modified by the effect of plasma
parameters, namely, $k$, $\alpha$, $\sigma$, $\sigma_i$, and
$\beta$. }

\item{It is clear from the expression of the phase speed ($V_p$)
that the phase speed greatly depends on the plasma parameters
$\alpha$ and $\beta$. From this observation, the phase speed of
HIA waves are found to be increased with the
decreasing value of $\alpha$ and $\beta$ (see Fig. 1).}

\item{The hump (compressive or positive) type HIA mK-dV solitons are found to
exist above the critical value ($k_c>1.48$). It is conspicuous
from the amplitude of positive potential mK-dV solitons that the
amplitude decreases with the increase of the heavy ion number
density $\alpha$ and the ion-fluid temperature $\sigma$ but
increases with the increasing values of $k$ (see Figs. 2-4).}

\begin{figure}[t!]
\centerline{\includegraphics[width=6.5cm]{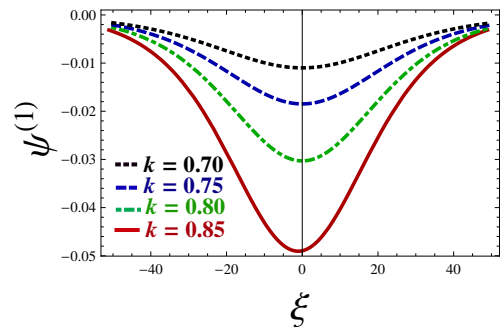}} \caption{The
variation of the negative potential GS solitons with $k$. The
other parameters are fixed at $\alpha=0.4$, $\sigma_i=1.4$,
$\beta=0.4$, and $U_{0}=0.01$.} \label{Fig.7}
\end{figure}

\begin{figure}[t!]
\centerline{\includegraphics[width=6.5cm]{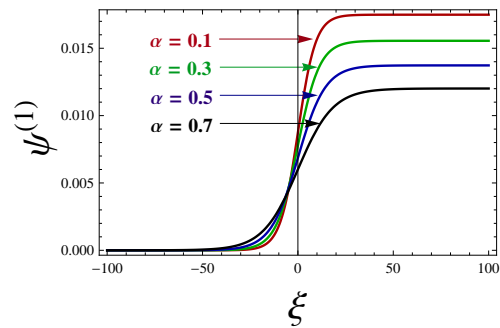}} \caption{The
variation of the positive potential DLs with $\alpha$ for
$k=1.4$, $\sigma_i=1.6$, $\beta=0.7$, and $U_{0}=0.01$.}
\label{Fig.8}
\end{figure}

\item{In our present attempt we have found the equation of HIA GS
which supports both the SWs and
DLs solutions by reason of containing both $\psi$-term of K-dV
and $\psi^2$-term of mK-dV equation. Therefore, the compressive
and rarefactive HIA GS solitons are observed to exist above the
critical value ($k_c>1.48$) and below the critical value
($k_c<1.48$) (see Figs. 5-7).}

\item{ The amplitude of the HIA GS increases with the increasing
values of spectral index (via $k$) and electron to heavy ion
number density (via $\beta$). Besides, the width of the negative
potential HIA GS becomes sharper with increases value of $k$ (see
Figs. 5-7).}

\item{It is found that only compressive HIA DLs are exist around
critical value ($k>k_d$). The amplitude of the positive potential
DLs are observed to increase with the decreasing values of ion to
heavy ion number density ratio (via $\alpha$) (see Fig. 8).}
\end{enumerate}

The principal features of HIA waves in an unmagnetized
collisionless EI plasma containing superthermal electrons,
Boltzmann distributed light ions, and adiabatic positively
charged inertial heavy ions have been theoretically analyzed. We
have numerically examined the effects of different plasma
parameters on the basic features (viz. polarity, amplitude, and
phase speed) of the HIA SWs as well as the adiabatic effects of
heavy ions and superthermality effects of electrons. The ranges
of plasma parameters used in our investigation ($\alpha=0.1-0.7$,
$\sigma_i=1.2-2$, $\beta=0.1-0.9$, $\sigma=0.01-0.2$)
\cite{Jilani2013}) which are relevant to astrophysical
\cite{Mendis1994,Goertz1989} and labrotory
\cite{Selwyn1993,Winter1998} plasma situations. It is obvious
from our present investigation that the effects of superthermal
electrons and adiabatic positively charged inertial heavy ions
greatly affect the fundamental properties of the HIA SWs.

Lastly, it is clarified from our present attempt that the results
that have been found can be useful to know the remarkable features
of HIA waves in collisionless unmagnetized plasmas where the
effect of superthermality of electrons become pronounced.

\textbf{Acknowledgments:} M. G. Shah, M. M. Rahman and M. R.
Hossen are greatly thankful to the Ministry of Science and
technology (Bangladesh) for rewarding the National Science and
Technology (NST) fellowship.


\end{document}